\documentclass[runningheads,a4paper]{llncs}

\usepackage{amssymb}
\usepackage{graphicx}

\usepackage{array}
\usepackage{booktabs}
\usepackage{multirow}
\usepackage{amsmath}

\usepackage{url}
\urldef{\mailsa}\path|{ym_zhang, luojcalways}@sjtu.edu.cn,|
\urldef{\mailsc}\path|{gao-xf, gchen}@cs.sjtu.edu.cn|    
\newcommand{\keywords}[1]{\par\addvspace\baselineskip
\noindent\keywordname\enspace\ignorespaces#1}

\begin{document}

\mainmatter  

\title{Which is better? A Modularized Evaluation for Topic
Popularity Prediction}

\titlerunning{Which is better? A Modularized Evaluation for Topic
Popularity Prediction}

%
%
\author{Yiming Zhang%
\and Jiacheng Luo\and Xiaofeng Gao\and Guihai Chen}
\authorrunning{Yiming Zhang et al.}

\institute{Shanghai Jiao Tong University, China\\
\mailsa\\
\mailsc\\
}

%
%

\maketitle

\begin{abstract}
Topic popularity prediction in social networks has drawn much attention recently. Various elegant models have been proposed for this issue. However, different datasets and evaluation metrics they use lead to low comparability. So far there is no unified scheme to evaluate them, making it difficult to select and compare models. We conduct a comprehensible survey, propose an evaluation scheme and apply it to existing methods. Our scheme consists of four modules: classification; qualitative evaluation on several metrics; quantitative experiment on real world data; final ranking with risk matrix and \textit{MinDis} to reflect performances under different scenarios. Furthermore, we analyze the efficiency and contribution of features used in feature oriented methods. The results show that feature oriented methods are more suitable for scenarios requiring high accuracy, while relation based methods have better consistency. Our work helps researchers compare and choose methods appropriately, and provides insights for further improvements.
\end{abstract}

\section{Introduction}\label{sec:intro}

Social network has become an indispensable part of our life, where users have access to posting and reading messages to and from the public. Everyday, in social networks, there are millions of messages posted about a wide range of topics. However, after a period of information dissemination, only a few prevailing topics will burst. Prediction of popularity of topics has meaningful impacts on many services and applications, such as marketing, advertisement, search engine queries and recommendation system~\cite{wu2017sequential,kong2014realtime}. News media utilizes online social networks to maximize their news visibility~\cite{Villi2016Recommend}. Hence, predicting the popularity of topics in social network is significant for both academia and industry.

Formally, we define the topic's popularity as the number of messages related to it~\cite{kong2014predicting} and the popularity fluctuates with time. The objective of topic popularity prediction problem is to detect topics and predict their popularities in the near future utilizing the obtainable information.
	
Various topic popularity prediction methods proposed by previous works have different properties such as accuracy and complexity, which leads to different suitability and feasibility under diverse scenarios. They all claim that they have great performance in the experiments. However, tested on different datasets and evaluation metrics, those methods have little comparability between each other, which makes it difficult to choose an appropriate one under specific scenarios. The objective of this paper is to perform a fair and strict evaluation of each topic popularity prediction method with a uniform evaluation scheme. 

We have collected the literatures about topic popularity prediction in social network published recently as completely as possible. After the survey, we, as the third party, develop a four-module evaluation scheme in order to make a convictive comparison to the topic popularity prediction methods comprehensively. We classify all the methods into two categories: \emph{feature oriented methods} and \emph{relation based methods} in the first module. In the second module, we carry out a qualitative evaluation representing the temporal, spatial, and platform universality. We also list some properties such as core algorithms and data preprocessing procedures, which fix their position among various problem situations. We further perform the scheme's third module: a quantitative experiment utilizing a dataset grabbed from Twitter. We test the methods based on the whole dataset without any optimization and give the accuracy evaluation which is objective and valuable to refer. In the fourth module, we rank the methods by $MinDis$ metric based on risk matrix~\cite{risk}. With various demands under different scenarios, we can use $ MinDis$ to evaluate models and choose an appropriate method.
		
Moreover, in feature oriented methods, features are the linchpin to ensure the accuracy. However, extracting features costs a lot, which requires to use efficient features to improve the accuracy and reduce the complexity at the same time. Therefore, we also give an evaluation of features' contribution. 

The contributions of our work are listed as follows:
\begin{itemize}
\item We conduct a comprehensive survey on topic popularity prediction in social network, which is shown in the qualitative evaluation section;
\item We propose a four-module evaluation scheme to fairly compare and rank existing or prospective methods under different scenarios;
\item We give an evaluation of features' contribution on this problem.
\end{itemize}
The paper is organized as follows. We first give some definitions in Section~\ref{sec:def}. In Section~\ref{section:overview}, we introduce our evaluation scheme. In Section~\ref{section:classification}, we classify the topic popularity prediction methods and give a qualitative evaluation of them. Section~\ref{section:quantitative} shows the quantitative comparison results of the methods. Section~\ref{section:ranking} talks about the final ranking module. Section~\ref{section:features} deals with the efficiency of the features using in machine learning models. In Section~\ref{section:conclusion} we give summaries and conclusions.

\section{Technical Preliminaries}\label{sec:def}

\textbf{Definitions:}
	The \textbf{social network} is defined as a graph, $SN=(U, E, M)$. $U$ is the set of the users (nodes) and $E$ is the set of the edges in the graph defined as ${(u_q,u_p)}$, which is weighted by times that $u_p$ interacts (mention or reply) with $u_q$. $M$ is the set of messages posted by $U$.
	A \textbf{topic} $h$ can be described as $(U^h,E^h,M^h)$, where $U^h$ have posted messages about $h$ , $M^h$, and $E^h$ are all the interactions at $h$. We divide the time $T$ into a series $T_1, T_2,..., T_t$. The \textbf{time series} of $h$ is $TS^h=TS_1^h, TS_2^h, ..., TS_t^h$. $TS_t^h$ is $(U^h_t,E^h_t,M^h_t)$ which means a part of $(U^h,E^h,M^h)$ where $M^h_t$ is the messages in $M^h$ posted in $T_t$ and $U^h_t$ only contains the users who post messages in $M^h_t$. Formally, the \textbf{popularity} $P$ of $h$ in time period $T_t$, $P^h_t$, is defined as $|M^h_t|$~\cite{kong2014predicting}.
\\
\textbf{Problem Statement:}
	Given a social network $SN$, the problem of topics popularity prediction is to predict $P$ of the topics in $T_{t+1}$ using the information, or features in $(U,E,M)$ before $T_t$. 
	In our evaluation, we specify the purpose of the prediction to be predicting whether a topic will be popular.

\section{The Evaluation Scheme}\label{section:overview}

\begin{figure}
			\vspace{-5mm}
			\centering
			\includegraphics[width=\linewidth]{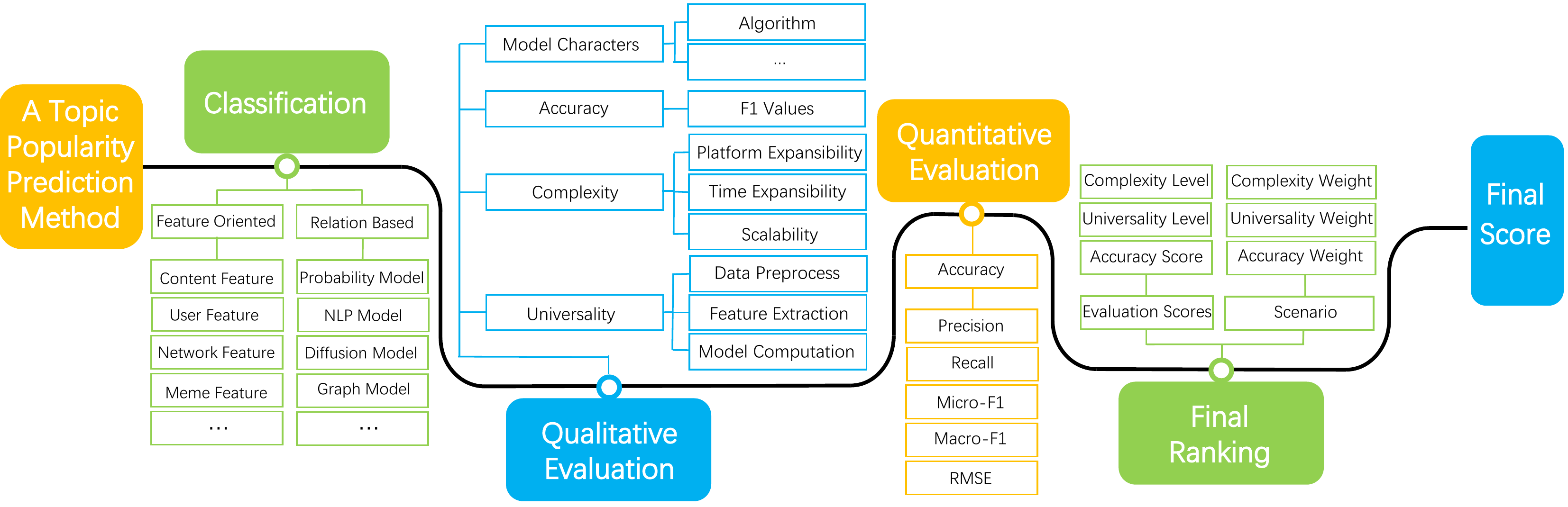}
			\caption{The Evaluation Scheme}
			\label{fig:scheme}
			\vspace{-5mm}
		\end{figure}

In this section, we elaborate our Topic Popularity Prediction Evaluation Scheme, which can be divided into 4 modules: \textit{Classification, Qualitative Evaluation, Quantitative Experiment}, and \textit{Final Ranking}. The scheme is shown in Figure~\ref{fig:scheme}.

\subsection{Classification and Qualitative Evaluation Overview}

According to our investigation, we classify the methods briefly into two categories: \textit{Feature oriented methods} and \textit{Relation based methods}. \textit{Feature oriented methods} use raw features and universal classification or regression models to predict. \textit{Relation based methods} aim at building a specific model by finding the mathematics relationship between the features and the popularity.
		
	Qualitative evaluation is performed to learn about the characteristics of each method. Three main metrics: accuracy\footnote{Accuracy is measured by F1 score.} ($a$), complexity (c), and universality ($u$), as well as other properties such as denoising processes and algorithm categories, are evaluated.
		
	We divide the three metrics into 3 levels and we provide a criteria table (Table~\ref{tab:papers information_criteria}) for them. These three metrics describe the synthetic performance of a method and we can use this criteria for all methods. 

\begin{table}
	\centering
	\tiny
	\renewcommand\arraystretch{1.5}
	\setlength{\tabcolsep}{0.18cm}
	\caption{\label{tab:papers information_criteria}Criteria on Qualitative Evaluation}
	\begin{tabular}{m{0.35in}|m{0.4in}m{2in}m{1.5in}}
		\toprule
		Level & Accuracy & Complexity\footnotemark[2] & Universality\\
		\midrule
		HIGH &$(60\%, 1)$ & If the complexity of preprocessing and extracting more than 20\% of the features is $O(n^2)$ or the main model's complexity is over $O(n^2)$. & Easy to transplant to other platforms (social network) and can deal with huge data stream in any time period\\
		MEDIUM & $(30\%, 60\%)$ &Few parts of the methods have a complexity over $O(n^2)$ but they are not in a absolute dominant role.& Have some limits to transplant to other platforms or can deal with common cases of data\\
		LOW & $(0, 30\%)$    & If most preprocessing procedures, feature extracting, and the main model can be done within $O(n)$. & Only be used in some specific situations or constricted to data scale\\
		\bottomrule	
	\end{tabular}
	\vspace{-3mm}
\end{table} 
\vspace{-2mm}
	\footnotetext[2]{$n$ is the number of tweets or users, and we only consider the number of traversal times in the complexity.}

\subsection{Quantitative Experiment Overview}
Qualitative evaluation gives out accuracy level of the methods. In this module  we evaluate precision more elaborately, excavate deep level information and achieve more precise and fairer evaluation.
	
Based on the unified big dataset from Twitter, we perform the experiments on topic popularity prediction methods, which makes the result comparable and gets rid of possible ``cherry-picking" problems. 

\subsection{Final Ranking Overview}
With the results of previous steps, we gain insights from risk management and employ risk matrix~\cite{risk} to generate the weights due to the nature that a model's final score should be considered by the category of probability against the category of importance. Then we rank models by presenting a novel metric, dubbed as \textbf{MinDis}.

We use $w_{a1}$, $w_{a2}$, $w_{rm}$, $w_c$, and $w_u$ to represent the importance weights of \textit{Macro-F1}, \textit{Micro-F1}, \textit{RMSE}, complexity, and universality.

\textbf{MinDis} is a metric proposed to describe how far a method is away from the ideal method according to the Euclidean Distance for a specific application scenario. It is formulated as:
			\begin{equation}
			MinDis=\sqrt{Dis_{F1_{mac}}+Dis_{F1_{mic}}+Dis_{rm}+Dis_{c}+Dis_{u}}
			\end{equation}

In the equation $Dis_{metrics}$ represents $w_{metrics}(E(metrics)-metrics)^2$. This shows the Euclidean Distance of a metrics between the evaluated method and the perfect one in theory. $MinDis$ can be used to rank all the methods under a specific scenario to select the most appropriate method.

\section{Classification and Qualitative Evaluation}\label{section:classification}

We first classify the methods and give a qualitative evaluation of them, which is aimed at comprehensively analyzing the method characteristics and giving an introduction of their merits and drawbacks. 
	
	\subsection{Feature Oriented Methods}
	In general, the topic popularity prediction problem for these methods can be divided into several parts: data preprocessing, denoising, feature extracting, and machine learning.
	For features are the key factors, in order to find efficient features, we briefly divide all the features further into six categories~\cite{kong2014predicting} (shown in Table~\ref{tab:papers information_machine learning}).
	
	In Table~\ref{tab:papers information_machine learning} we list all the metrics used to investigate the feature oriented methods in our evaluation scheme. Some methods treat hashtags as topics and utilize directly obtained features to train a classifier~\cite{ma2013predicting,kong2014predicting}, while other methods pre-process the features to detect the topics and then train the machine learning algorithms~\cite{chen2013emerging}. The metrics in the evaluation scheme are described below:\\
\textbf{Accuracy, Complexity, and Universality:} Refer to Table~\ref{tab:papers information_criteria}.
		\\
\textbf{F. Categories:} Features impact the accuracy and efficiency directly. In Table~\ref{tab:papers information_machine learning}, we list the categories of features a method adopts and the size of the feature space (\textbf{Number of Features}). Due to the limited space, we use \textbf{1} - \textbf{6} to denote six kinds of features: \textbf{Content}, \textbf{User}, \textbf{Network}, \textbf{Meme}, \textbf{Hashtag}, and \textbf{TimeSeries} respectively. We further evaluate feature effectiveness in Section~\ref{section:features}.  
		\\
\textbf{Denoise:} Denoising is an influential procedure to promote the performance of prediction.  Some methods generate an equal internal time series (TS) to obtain information in early time intervals. Some use \emph{critical drop} point to extract potential burst keywords. Others detect emerging topics (DET), and predict whether they will burst.

\begin{table}
		\centering
		\tiny
		\renewcommand\arraystretch{1.5}
		
		\caption{\label{tab:papers information_machine learning}Qualitative Evaluation of Feature Oriented Methods}
		\begin{tabular}{m{0.8cm}m{0.55cm}<{\centering}m{0.8cm}<{\centering}m{0.9cm}<{\centering}m{1.1cm}<{\centering}m{1.3cm}<{\centering} m{0.6em}<{\centering}m{0.6em}<{\centering}m{0.6em}<{\centering}m{0.6em}<{\centering}m{0.6em}<{\centering}m{0.6em}<{\centering} m{0.8cm}<{\centering}m{1.3cm}p{0.6cm}<{\centering}}
			\toprule
			\multirow{2}{*}{Method} & \multirow{2}{*}{Paper}& \multirow{2}{*}{Publish} & \multirow{2}{*}{Accuracy} & \multirow{2}{*}{Complexity} & \multirow{2}{*}{Universality} &\multicolumn{6}{c}{F. Categories} & \multirow{2}{0.7cm}{Number of Features} & \multirow{2}{*}{Algorithm} & \multirow{2}{*}{Denoise} \\
			\cmidrule{7-12}
			& &  &  &  & & 1 & 2 & 3 & 4 & 5 & 6 &  &  &  \\
			\midrule
			F-\uppercase\expandafter{\romannumeral1} & \cite{ma2013predicting}{\tiny} & AS-IS\&T & Medium & Medium & High & * & * & * & * & * &    & 18& Classification & TS  \\
			
			F-\uppercase\expandafter{\romannumeral2} & \cite{kong2014predicting}{\tiny} & SIGIR & Low & High & High & * & * & * & * & * & *  & 26 & Classification & TS \\
			
			F-\uppercase\expandafter{\romannumeral3} & \cite{du2011microblog}{\tiny} & ITAIC & High & High & Medium &   & * & * &   &   &    & 6&  Unsupervised Learning &  CD\\
			
			F-\uppercase\expandafter{\romannumeral4} & \cite{ishikawa2012hot}{\tiny} & ARCS & - & High & Medium & * &  &  & * & * &   & 3& Clustering &  -  \\
			
			F-\uppercase\expandafter{\romannumeral5} & \cite{chen2013emerging}{\tiny} & SIGIR & High & High & Medium & * &  &  & * &  &   & 6 & Co-training \& semi-supervised &  DET \\
			
			F-\uppercase\expandafter{\romannumeral6} & \cite{becker2011beyond}{\tiny} & ICWSM & High & High & Medium & * &  & * &  & * & * & Many& Classification &  -  \\
			
			F-\uppercase\expandafter{\romannumeral7} & \cite{wu2017sequential}{\tiny} & IJCAI & - & High & Low & * & * &  &  &  &  & - & DTCN & - \\
			
			F-\uppercase\expandafter{\romannumeral8} & \cite{liu2014trending}{\tiny} & IC-BNMT & High & Low & High & * & * & * & * &  &  & 8 & SVM & - \\

			\bottomrule	
		\end{tabular}
	\end{table}

\subsection{Relation Based Methods}
Different from feature oriented methods, relation based methods regard the social network as heterogeneous distributions.

Geographical records of social networks work as sensors to find regions' interests. Thus,~\cite{yin2011geographical} proposed a geographical method, Latent Geographical Topic Analysis (LGTA), which combines contents and locations. A much more popular type of methods is hashtag-based\cite{cui2012discover}. Performances of these methods are more stable than others due to the empirical studies of burst hashtags patterns. Many other models are proposed based on Natural Language Processing (NLP), probability theory, graph theory, and so on. Clustering is another interesting approach. ~\cite{hoang2017gpop} clustered users into groups and predicted the group-level popularity which improves the accuracy.

The evaluation results are listed in Table~\ref{tab:papers information_modeling}. The chosen metrics in the evaluation platform can be described as follows:\\
\textbf{Accuracy}, \textbf{Complexity}, and \textbf{Universality} are the same as above.
		\\
		\textbf{Recognition Time:} This metric depicts how long a method can predict, such as R-\uppercase\expandafter{\romannumeral4} can predict a topic two days prior to its burst on the civil unrest datasets.
		\\
		\textbf{Data Preprocessing:} Some methods preprocess data before prediction, thus have high dependences on how well data are preprocessed.
		\\
		\textbf{Field:} We summarize the field of knowledge the methods base on. Geographical methods are accessible in most platforms and languages, with a relative low accuracy. Methods based on graph theory and NLP can detect hot topics much earlier than others. However, they have a heavy dependence on preprocessing procedure.

\begin{table}
		\vspace{-3mm}
		\centering
		\setlength{\tabcolsep}{0.13cm}
		\tiny
		\renewcommand\arraystretch{1.5}
		\caption{\label{tab:papers information_modeling}Qualitative Evaluation of Relation Based Methods}
		\begin{tabular}{m{0.55cm}<{\centering} m{0.5cm}<{\centering} m{1.5cm}<{\centering} m{1.4cm} m{0.7cm}  m{1.0cm} m{1.0cm} m{1.0cm} m{1cm}<{\centering} m{1.0cm}}
			\toprule
			Method & Paper & Published in & Model & Accuracy & Complexity & Recognition Time & Universality & Data Preprocessing & Field\\
			\midrule
			R-\uppercase\expandafter{\romannumeral1} & \cite{cui2012discover} & CIKM & IMA & Medium & Low & - & High &   & Probability (hashtag)\\
			
			R-\uppercase\expandafter{\romannumeral2} & \cite{deng2015effectively} & AAAI & PreWHether
			& High & Low & Medium & Medium &   & Probability\\
			
			R-\uppercase\expandafter{\romannumeral3} & \cite{zhu2014prerecognition} & Sci. W. J. & Prerecognition Model
			& Medium & Medium & High & High & * & NLP\\
			
			R-\uppercase\expandafter{\romannumeral4} & \cite{chen2014non} & SIGKDD & NPHGS & High & High & High & Medium & * & Graph Theory\\
			
			R-\uppercase\expandafter{\romannumeral5} & \cite{yin2011geographical} & IW3C2 & LGTA & Low & Low & Low & High &  & Geographical\\
			
			R-\uppercase\expandafter{\romannumeral6} & \cite{sakaki2010earthquake} & WWW & Temporal and Spatial Model &  Medium & Medium & Medium & Medium &  & Geographical\\
			
			R-\uppercase\expandafter{\romannumeral7} & \cite{amodeo2011hybrid} & CIKM & Hybrid Model  & High & Medium & Low & High & * & Probability\\
			
			R-\uppercase\expandafter{\romannumeral8} & \cite{zhang2015event} & Neurocomputing & Diffusion Model  &  Medium & Medium &  Medium & High &  & Graph Theory\\
			
			R-\uppercase\expandafter{\romannumeral9} & \cite{chen2013latent} & NIPS & Latent Source Model &  High & Low &  Low & High &  & Probability\\
			
			R-\uppercase\expandafter{\romannumeral10} & \cite{matsubara2017nonlinear} & TWEB & SPIKEM & High & Medium & - & High &  & Probabilistic \\
			
			R-\uppercase\expandafter{\romannumeral11} & \cite{hoang2017gpop} & WWW & GPOP & High & Medium & High & High & * & Tensor decomposition \\
            
            R-\uppercase\expandafter{\romannumeral12} & \cite{wang2017predicting} & JIIS & EdgeMRF & - & Medium & - & High &  &  MRF\\
		\bottomrule
		\end{tabular}
		\vspace{-2mm}
	\end{table}	

	\section{Quantitative Experiment}\label{section:quantitative}
	We perform a quantitative experiment based on a real world dataset to guarantee the fairness. We choose five methods to evaluate: F-\uppercase\expandafter{\romannumeral1}, F-\uppercase\expandafter{\romannumeral2}, R-\uppercase\expandafter{\romannumeral1}, R-\uppercase\expandafter{\romannumeral2} and R-\uppercase\expandafter{\romannumeral3}, for they are all proposed  recently, contain relatively complete feature spaces, and have great universality.
	
	\subsection{Introduction of the Experiments }
	Our dataset is grabbed from Twitter containing about 2 million tweets from August 1, 2015 to January 31, 2016. After filtering, all the data we used are tweets post in English because different languages may need different tokenization methods. Table~\ref{tab:info.database} shows the detailed statistics about the dataset. In our evaluation, the time period is set to be $1$ day. 
    	\vspace{-2mm}
		\begin{table}[htbp]
			\centering
			\setlength{\tabcolsep}{0.05in}
			\renewcommand\arraystretch{0.82}
			\caption{\label{tab:info.database}Detailed Information about the Dataset}
			\begin{tabular}{p{1.3in}p{0.6in}p{2in}p{0.4in}}
				
				\toprule
				Statistic & Value & Statistic & Value\\
				\midrule
				Number of Tweets & 2.5 M & Number of Followers & 4.5 M\\
				Number of Users & 1.5 M & Number of Tweets with Hashtag & 35K\\ 
				Number of Retweet & 0.7 M & Fraction of Tweets with Hashtag & 1.4\%\\
				
				\bottomrule
				
			\end{tabular}
		\end{table}
        \vspace{-2mm}
	
	We conduct 10-fold cross validation on all the data to train the methods using machine learning models. All the features are extracted without any unfair optimization. We choose 5 metrics: \textit{Precision}, \textit{Recall}, \textit{Macro-F1}, \textit{Micro-F1}, and \textit{RMSE} to measure their performance. $F\emph{1}$ is an metric that synthesizes precision and recall. \textit{Macro-F1} is impacted mainly by the prediction accuracy of rare classes (popular topics).  \textit{Micro-F1} is similar to \textit{Macro-F1}, but it is impacted mainly by the prediction accuracy of common classes (non-popular topics). We employ \textit{RMSE} to uncover integral constancy of the methods

\subsection{Results of the Quantitative Experiment}
\begin{figure}
		\begin{minipage}[t]{0.5\linewidth}
			\vspace{30pt}
			\centering
			\includegraphics[width=1.02\linewidth]{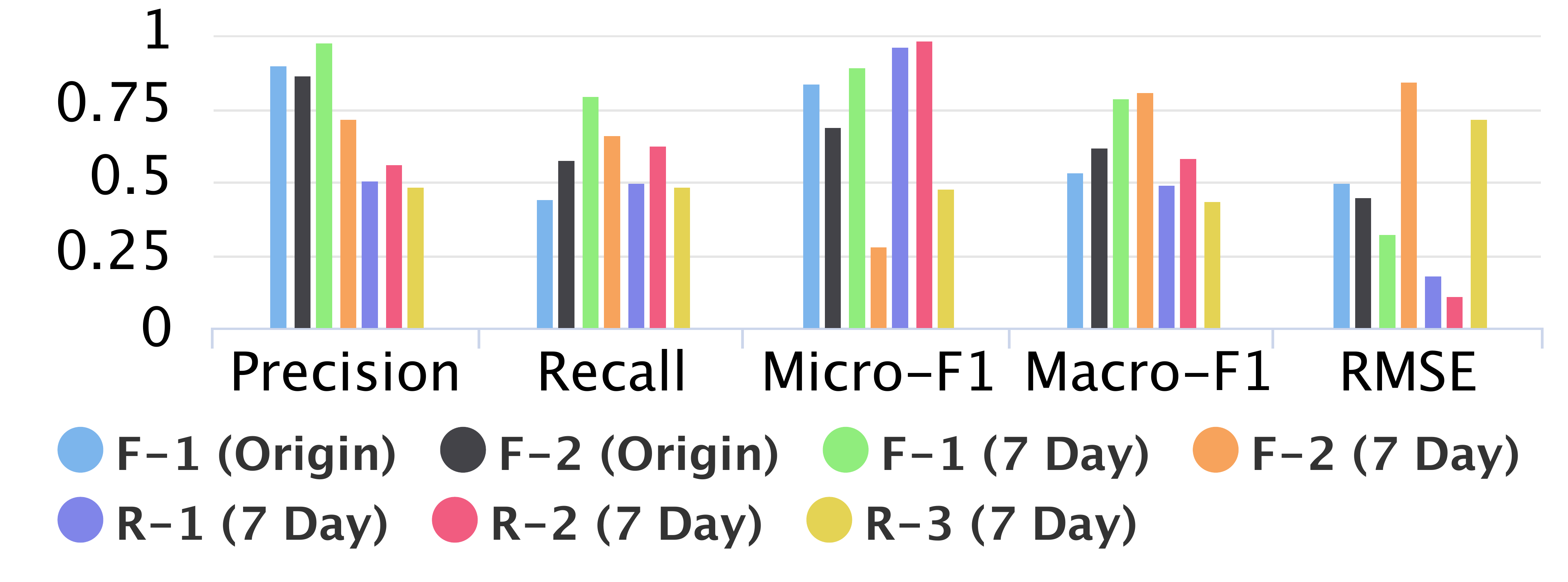}
			\caption{Five Index Scores of Each Method}
			\label{fig:ML-M}
		\end{minipage}
		\hfill
        \begin{minipage}[t]{0.5\linewidth}
        \makeatletter\def\@captype{table}\makeatother
			\vspace{0pt}
			\centering
			\fontsize{6.6pt}{7pt}\selectfont
			\caption{\label{tab:Paper Evaluation}Five Index Scores of Each Method}
			\begin{tabular}{m{1.0cm}p{1.1cm}<{\centering}p{0.8cm}<{\centering}p{0.35in}<{\centering}p{0.35in}<{\centering}p{0.35in}<{\centering}}
				
				\toprule
				Method & Precision & Recall & Macro-F1 & Micro-F1 & RMSE \\
				\midrule
				F-\uppercase\expandafter{\romannumeral1} (Origin) & 0.9064 & 0.4421 & 0.5367 & 0.8381 & 0.4983 \\
				F-\uppercase\expandafter{\romannumeral2} (Origin) & 0.8671 & 0.5798 & 0.6194 & 0.6915 & 0.4554 \\
				F-\uppercase\expandafter{\romannumeral1} \hspace{0.9em} (7 Day) & \textbf{0.9789} & \textbf{0.8000} & 0.7889 & 0.8947 & 0.3244 \\
				F-\uppercase\expandafter{\romannumeral2} \hspace{0.5em} (7 Day) & 0.7619& 0.6667 & \textbf{0.8148} & 0.2857 & 0.8452 \\
				R-\uppercase\expandafter{\romannumeral1} \hspace{0.9em} (7 Day) & 0.5070& 0.5005 & 0.4942 & 0.9668 & 0.1822 \\
				R-\uppercase\expandafter{\romannumeral2} \hspace{0.5em} (7 Day) & 0.5627& 0.6290 & 0.5839 & \textbf{0.9879} & \textbf{0.1100} \\
				R-\uppercase\expandafter{\romannumeral3} \hspace{0.5em} (7 Day) & 0.4856 & 0.4907 & 0.4358 & 0.4800 & 0.7211 \\
				\bottomrule
			\end{tabular}
			
		\end{minipage}
	\end{figure}

	We choose the data in 7 days to fairly compare both two kinds of methods. Feature oriented methods are further experimented with the half-year data, dubbed as (Origin). In Table~\ref{tab:Paper Evaluation}, we list the accuracy of each method and we compare them in Figure~\ref{fig:ML-M}.

\textbf{Comparison between Feature Oriented Methods and Relation Based Methods:}
	From Figure~\ref{fig:ML-M}, 
	we can find that feature oriented methods perform better than relation based methods in precision, recall, and Macro-F1. This meets expectations, for feature oriented methods take more factors into account and the process
	 of training with big data can give a more exact relationship between the factors and  popularity. While \textit{RMSE} and \textit{Micro-F1} scores shows that feature oriented methods have a lower performance in prediction of non-popular topics, which may be because the model is over-fitting.
	
\textbf{Comparison within Feature Oriented Methods:}
	From Figure~\ref{fig:ML-M}, we can see that when predicting for only 7 days,  F-\uppercase\expandafter{\romannumeral1} with a smaller feature space has a better performance. It gives an expression to the importance of feature choice and the process to get rid of noise. For denoising, in F-\uppercase\expandafter{\romannumeral1}, only information in time interval $t$ is used to denoise. While in F-\uppercase\expandafter{\romannumeral2}, time interval $t-4, t-3,..., t-1$ are also used. It is possible that F-\uppercase\expandafter{\romannumeral1} does a better job in denoising in this short period prediction because the information used for denoising in F-\uppercase\expandafter{\romannumeral1} is closer to the predicted time period and can reflect the features more accurately.
\\
F-\uppercase\expandafter{\romannumeral1} (7 Day) has a higher accuracy than F-\uppercase\expandafter{\romannumeral1} (Origin). However, F-\uppercase\expandafter{\romannumeral2} (7 Day) is worse than F-\uppercase\expandafter{\romannumeral2} (Origin). This shows  F-\uppercase\expandafter{\romannumeral1} is more suitable in short period prediction than F-\uppercase\expandafter{\romannumeral2}, while F-\uppercase\expandafter{\romannumeral2} is more suitable in predicting rare class in both situations for its Macro-F1 value is higher. In long period prediction, both of them have great accuracies.

\textbf{Comparison within Relation Based Methods:}
	From Figure~\ref{fig:ML-M} we find that  R-\uppercase\expandafter{\romannumeral2} consistently achieves better performance than the other two relation based methods. The three latent features: \textit{Sum, Average Rate of Change,} and \textit{Standard Deviation} help the model to detect hot topic patterns in a universal way.
	\\
However,  R-\uppercase\expandafter{\romannumeral1} uses three features detected from experiential studies. Setting a class of miscellaneous does not always perform well because their patterns are not constant and will have different performances in different datasets. In our experiments, R-\uppercase\expandafter{\romannumeral1} is too sensitive to find enough hot topics. More effective features should be implemented to better classify patterns of events. 
	\\
R-\uppercase\expandafter{\romannumeral3} is a language-related solution. Its performance displayed in Figure~\ref{fig:ML-M} is quite different from the other two, especially on Micro-F1 and RMSE. Micro-F1 proves that NLP methods are apt to predict with a large scale of data, or "rich" data. For its experiments are based on microblogging data, we find it defective to process Twitter data and this method highly depends on pretreatment of posts, and users should manually reprocess the clustering results.  A high RMSE reveals its precariousness.

\section{Final Ranking}\label{section:ranking}

\vspace{-3mm}
	With the evaluation results above, we rank the methods by \textit{MinDis} metric to fit different demands. We implement the ranking steps under 4 classical scenarios respectively.
	\\
\textbf{\uppercase\expandafter{\romannumeral1}}: This scenario is under a balance requirement, where the importances of every metric are the same. 
\\
\textbf{\uppercase\expandafter{\romannumeral2}}: This scenario is under a complexity oriented requirement which is widely used in real time prediction. Low complexity and high universality are required to deal with large flow data stream from different platform.
\\
\textbf{\uppercase\expandafter{\romannumeral3}}: Accuracy is heavily weighted and Macro-F1 has the biggest weight, which shows that the precise prediction of rare classes is important under this scenario. This situation conforms to recommendation systems' requirement.
\\
\textbf{\uppercase\expandafter{\romannumeral4}}: We highly consider consistency with the goal of choosing a method with consistent performance. 

\vspace{-8mm}
\begin{figure}
            \begin{minipage}[t]{0.42\linewidth}
            \makeatletter\def\@captype{table}\makeatother
            \centering
            \fontsize{6.5pt}{8.1pt}\selectfont
			\renewcommand\arraystretch{1}
            \caption{\label{tab:riskmatrix }Risk Matrix of Scenario \uppercase\expandafter{\romannumeral3}}
            \begin{tabular}{m{1cm}<{\centering} | m{1.1cm}<{\centering} | m{1cm}<{\centering} | m{1cm}<{\centering} | m{1.3cm}<{\centering}}
            \toprule
                & Negligible & Marginal & Critical & Catastrophic \\ 
            \hline
            Certain & & & & \\ 
            \hline
            Likely & & $c$ & & \\ 
            \hline
            Possible & & $u$ & $F1_{mic}$, $rmse$ & $F1_{mac}$, \\ 
            \hline
            Unlikely & & & & \\ 
            \hline
            Rare & & & & \\ 
            \bottomrule
            \end{tabular}
            \end{minipage}
            \hfill
            \begin{minipage} [t]{0.5\linewidth}
            \makeatletter\def\@captype{table}\makeatother
			\centering
            \fontsize{6.5pt}{8.1pt}\selectfont
			\renewcommand\arraystretch{1}
			\caption{\label{tab:Weight} Weights under Each Scenario }
			\begin{tabular}{m{1cm}<{\centering}  m{0.7cm}<{\centering}m{0.7cm}<{\centering}m{0.7cm}<{\centering}m{0.7cm}<{\centering}m{0.7cm}<{\centering}}

				\toprule
				Scenarios & $w_c$ & $w_u$ & $w_{a1}$ & $w_{a2}$ & $w_{rm}$\\
				\midrule
				\uppercase\expandafter{\romannumeral1}	& 0.200 	& 0.200 	& 0.200 	& 0.200 	& 0.200  \\
				\uppercase\expandafter{\romannumeral2} 	& 0.286 	& 0.285 	& 0.143 	& 0.143 	& 0.143 \\
				\uppercase\expandafter{\romannumeral3} 	& 0.182 	& 0.091 	& 0.363		& 0.182 	& 0.182  \\
				\uppercase\expandafter{\romannumeral4}	& 0.125 &	0.125&	0.125&	0.125&	0.5\\
				\bottomrule
				
			\end{tabular}

            \end{minipage}
			
	\end{figure}
    \vspace{-6mm}
		
	We adopt risk matrix to determine the weight of metrics under different scenarios. The columns show different levels of severity if the metrics are too bad while the rows show how possible and frequent each good metric can be achieved in all methods. Each level has an interval of 20\%. According to our survey, 50\% methods have low accuracy, thus, the likelihood of Micro-F1 and Macro-F1 is Possible. Similarly, the likelihood of high complexity, low universality and high RMSE are likely, possible and possible respectively. Table \ref{tab:riskmatrix } shows the risk matrix of scenario \textbf{\uppercase\expandafter{\romannumeral3}}. The weights of metrics under it can thus be specified as in Table \ref{tab:Weight}. We do the same for scenario \textbf{\uppercase\expandafter{\romannumeral2}}, \textbf{\uppercase\expandafter{\romannumeral3}} and \textbf{\uppercase\expandafter{\romannumeral4}}. Users can make up their own scenarios.
	
	We use the weights and previous evaluation results to calculate \emph{MinDis}. The values of low, medium, and high level are 0.4, 0.5 and 0.6 because they can keep the relatively balance relationship among the metrics and guarantee that there is no one metric overwhelming in the final results. The results are listed in Table~\ref{tab:Ranking}. F-\uppercase\expandafter{\romannumeral1} (7 Day) outperforms others under three scenarios while R-\uppercase\expandafter{\romannumeral2} (7 Day) stands out under scenario  \uppercase\expandafter{\romannumeral4} showing a better consistency.
.
	
\begin{table}
		\vspace{-6mm}
		\centering
		\renewcommand\arraystretch{1}
		\scriptsize
		\caption{\label{tab:Ranking} Final Ranking of All the Methods in Each Scenario }
		\begin{tabular}{m{1.2cm}<{\centering}  m{1.3cm}<{\centering}m{1.3cm}<{\centering}m{1.3cm}<{\centering}m{1.3cm}<{\centering}m{1.3cm}<{\centering}m{1.3cm}<{\centering}m{1.4cm}<{\centering}}

			\toprule
			Scenarios & F-\uppercase\expandafter{\romannumeral1} (Origin) & F-\uppercase\expandafter{\romannumeral2} (Origin)&  F-\uppercase\expandafter{\romannumeral1} (7 Day) & F-\uppercase\expandafter{\romannumeral2} (7 Day) & R-\uppercase\expandafter{\romannumeral1} (7 Day) & R-\uppercase\expandafter{\romannumeral2} (7 Day) & R-\uppercase\expandafter{\romannumeral3} (7 Day)\\
			\midrule
			
			   \uppercase\expandafter{\romannumeral1} & 0.3160 &
0.2991 & \textbf{0.1848} & 0.5018 & 0.2471 & 0.2170 & 0.4730 \\
			    \uppercase\expandafter{\romannumeral2} & 0.2697 &
0.2528 & \textbf{0.1608} & 0.4241 & 0.2221 & 0.2019 & 0.4016 \\
			    \uppercase\expandafter{\romannumeral3} & 0.3603 & 0.3282 & \textbf{0.1979} & 0.4849 & 0.3194 & 0.2709 & 0.5112 \\
			    \uppercase\expandafter{\romannumeral4} & 0.3943 & 0.3656 & 0.2466 & 0.6521 & 0.2031 & \textbf{0.1842} & 0.5786\\
			\bottomrule
			
		\end{tabular}
	\end{table}
	\vspace{-10mm}

\section{Evaluation of Effectiveness of the Features}\label{section:features}
	 In this section, we analyze the features widely used in this problem, and evaluate their efficiency based on numerical experiments in order to choose appropriate features. LibSVM is used as the machine learning model. We train the model using the whole dataset, and then 7-day data. 10-fold cross validation is used to separate training and testing data.

	Altogether we take 34 features into account. The features are classified into six categories~\cite{kong2014predicting}. We list all the features in Table~\ref{tab:Features information} with some qualitative analyses: $Complexity$ which means the complexity to extract the feature, $Temporal$ representing whether this feature varies with time, and $Graphic$ telling us whether this feature describes graph $(U^h_t, E^h_t)$. 

	The features are described as follows:
		\begin{table} [htbp]
			\scriptsize
			\centering
			\setlength{\tabcolsep}{0.15cm}
			\renewcommand\arraystretch{0.85}
			\caption{\label{tab:Features information}Features Used in Feature Oriented Methods}
			\begin{tabular}{p{0.5cm}p{1.5cm}p{6.5cm}p{1.5cm}<{\centering}p{0.8cm}<{\centering}p{0.8cm}<{\centering}}
				\toprule
				
				No. & Name & Description & Complexity & Temporal& Graphic \\
				\midrule
				$ F_{c1} $ & NumEmo & The total number of emoticons included in $T^{h}_{t}$ & Medium & * & \\
				$F_{c2}$ & NumSpeSig & The total number of special signals like ``!!!!'' or ``Goooood!'' included in $T^{h}_{t}$ & Medium & * & \\
				$ F_{c3} $ & SentiOHash & The value of positive and negative points calculated by SentiStrength model& Medium &  & \\
				$ F_{c4} $ & Topics & 20-Dimension topic distribution vector derived from $T^{h}_{t}$ using LDA model & High &  & \\
				$F_{u1}$ & ActOUser & Average activities of $U^{h}_{t}$ gained by PageRank~\cite{Page1998The} & Medium & * & *\\
				$F_{u2}$ & MaxOF & The max number of followers of users in $U^{h}_{t}$ & Medium & * & *\\
				$F_{u3}$ & AvOF & Average number of followers of users in $U^{h}_{t}$ & Medium & * & *\\
				
				$F_{h1}$ & Length & The length of the hashtag & Low &  & \\
				$F_{h2}$ & multiFreq & Fraction of tweets having more than one hashtags & Medium & *  & \\
				$F_{h3}$ & Clarity & Kullback Leibler divergence of word distribution between $T^{h}_{t}$ and tweets collection T & High &  & \\
				$F_{h4}$ & ExtClar & The extension of $F_{h3}$ & Quite High &  & \\
				$F_{h5}$ & NumInHash & Whether a number is contained in the hashtag & Low &  & \\
				$F_{h6}$ & NumOWord & The number of words in a hashtag & Low &  & \\
				$F_{n1}$ & Degree & The average degree of nodes in $G^{h}_{t}$ & High & * & *\\
				$F_{n2}$ & Density & The density of $G^{h}_{t}$ & High & * & *\\
				$F_{n3}$ & Order & The order of $G^{h}_{t}$ & High & * & *\\
				$F_{n4}$ & EntrODD & The entropy of degree distribution of $G^{h}_{t}$ & High & * & *\\
				$F_{n5}$ & NumOBUser & The number of border users  of $G^{h}_{t}$ & Quite High & * & *\\
				$F_{n6}$ & ExpVec & Exposure vector of border users & Quite High &  * & *\\
				$F_{n7}$ & CompFrac & Ratio between number of connected components and number of nodes in $G^{h}_{t}$ & High & * & *\\
				$F_{n8}$ & Weight & The average weight of edges in $G^{h}_{t}$ & High & * & *\\
				$F_{n9}$ & TriFrac & Fraction of users forming triangles in $G^{h}_{t}$ & High & * & *\\
				
				$F_{m1}$ & NumOUser & The number of users in the $U^{h}_{t}$ & Medium & *  & \\
				$F_{m2}$ & FracOUser & The fraction of users who have posted tweets with hashtag $h$ in the $G_{t}$ & High & *  & \\
				$F_{m3,4}$ & NumO@, FracO@ & The Number and feaction of tweets in which the author has @ someone else in the $T^{h}_{t}$ & Medium & * & \\
				$F_{m5,6}$ & NumORT, FracORT & The Number and fraction of retweets in the $T^{h}_{t}$ & Medium & * & \\
				$F_{m7}$ & NumOT & The number of tweet in the $T^{h}_{t}$ & Medium & * & \\
				$F_{m8}$ & FracOURL & The fraction of tweet having URL in it in the $T^{h}_{t}$ & Medium & * & \\
				$F_{t1}$ & Mn & The mean value of the time series & High & * & \\
				$F_{t2}$ & MnD & The standard deviation value of the time series & High & * & \\
				$F_{t3}$ & Sd & The mean value of the absolute first-order derivative of time series & High & * & \\
				$F_{t4}$ & SdD & The standard deviation value of the absolute first-order derivative of time series & High & * & \\
				\bottomrule
				
			\end{tabular}
		\end{table}	        
\\   
\textbf{Content Features $F_{c}$:} They reflect the meaning and sentiment of posts. Previous studies show that negative topics are more likely to be popular~\cite{thelwall2011sentiment}. Therefore, the features describing sentiment ($F_{c1}, F_{c2}, F_{c3}$) are taken into account.
\\		
\textbf{User Features $F_{u}$:} Because a tweet posted by a more active user is more likely to be popular, $F_{u1}$ is listed in. $F_{u2}$ and $F_{u3}$ are extracted to reflect the number of followers who affect the dissemination of tweets.
\\		
\textbf{Hashtag Features $F_{h}$:} Hashtag Features show the number of words ($F_{h6}$), clarity ($F_{h3}$), and some other information. For instance, ``livemusic'' and ``Epl'' are two hashtags. ``livemusic'' has two words and a clearer meaning. 	
\\	
\textbf{Network Features $F_{n}$:} The graph $(U^h_t,E^h_t)$ is described by network features. Border users represent those who is not in  $U^h_t$ but follows the users in $U^h_t$.  $F_{n6}$ is a 15 dimension vector and dimension $D_i$ denotes that the number of border users following $i$ users in $ U^h_t $ is $D_i$. A three nodes cycle in the graph is a triangle and let the set of triangles be $\varDelta^h_{t}$. $F_{m2}$ is $|\varDelta^h_t| \left. \right/ (|U^h_t|(|U^h_t|-1)(|U^h_t|-2)\left. \right/ 3)$. Let $ C^h_t $ stands for the set of disconnected parts of the graph. $F_{n7}$ is defined as $| C^h_t| \left. \right/ |U^h_t |$.
	\\	
\textbf{Meme Features $F_{m}$:} Meme Features are mainly about  ``numbers'' and ``fractions'' in tweets such as the number and fraction of tweets with ``@''.
\\	
\textbf{Time Series Features $F_{t}$:} Time series can be described by the polynomial fitting curve and the absolute first-order derivative of the fitting curve. Their mean value and standard deviation value  are chosen as features.

	Some of these methods are multidimensional such as Topics ($F_{c4}$) whose dimension is 20, so the whole feature space has 68 dimensions. We remove each feature from the feature space~\cite{gao2014effective}, use the new feature space to train the classifier respectively, and rank all the removed features by its relative contribution (RC), which is:
	\vspace{-2mm}
	$$RC=-1000(A_i-A_s)\vspace{-2mm}$$
	$A_i$ is the average value of \textit{Macro-F1} and \textit{Micro-F1} when $F_i$ is removed from the feature space and $A_s$ is the standard average value with the whole feature space.
	
\begin{figure}
		\vspace{-3mm}
		\makeatletter\def\@captype{figure}\makeatother
        \begin{minipage}[t]{0.5\linewidth}
			\vspace{12em}
			\centering
			\includegraphics[width=\linewidth]{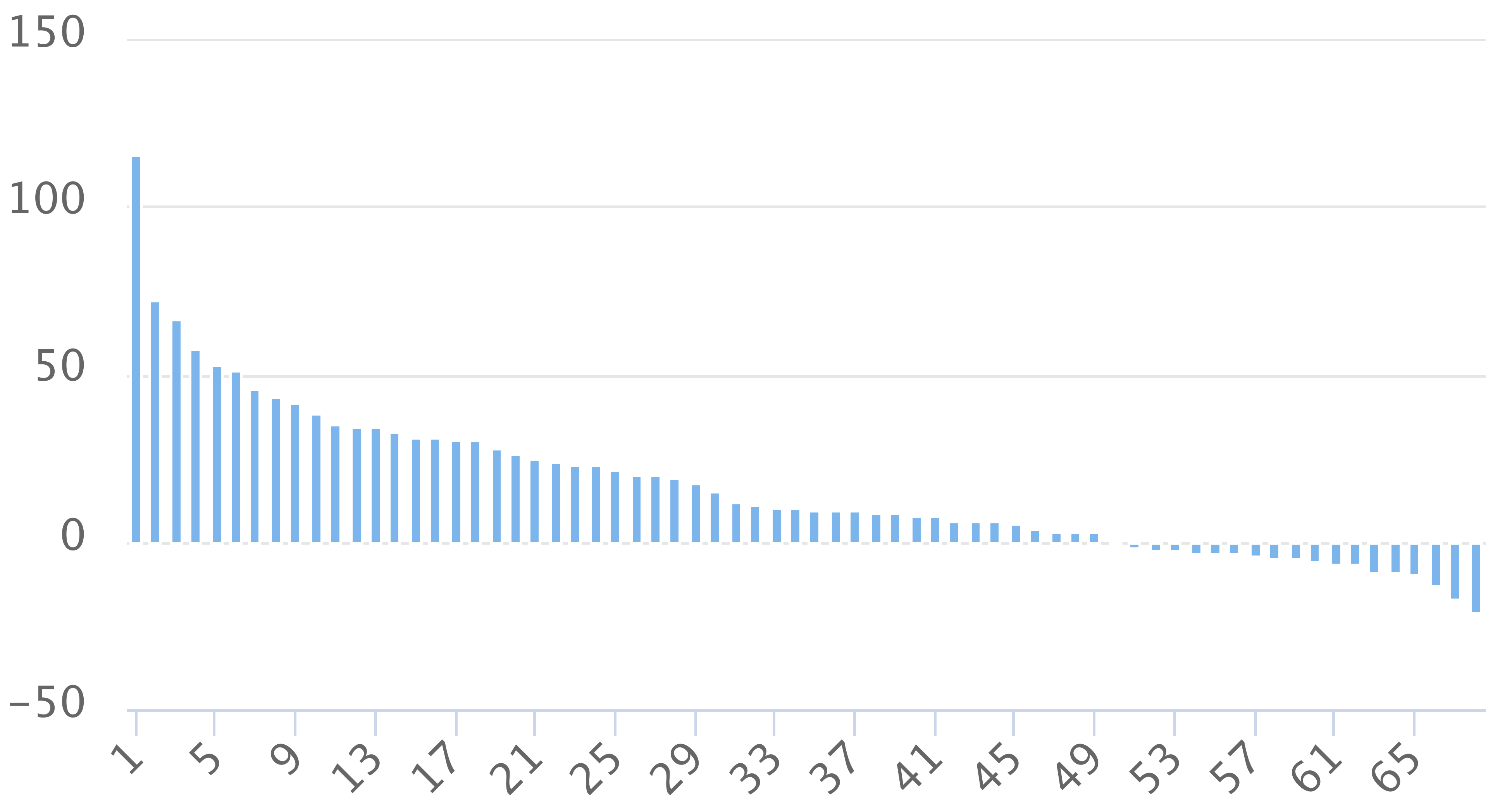}
			\caption{The Relative Contribution of the Feature}
			\label{fig:features}
		\end{minipage}
		\hfill
		\begin{minipage}[t]{0.5\linewidth}
			\makeatletter\def\@captype{table}\makeatother
			\vspace{0pt}
			\centering
			\renewcommand\arraystretch{0.9}
			\tiny
			\caption{\label{tab:Feature Rank}The 15 Most Efficient Features and the 15 Least Ones }
			\begin{tabular}{m{0.5cm}m{2.0cm}m{0.5cm}p{2.3cm}}

				\toprule
				Rank & Feature & Rank & Feature\\
				\midrule
				1 & $F_{n5}$: Number of Border User & 54 & $F_{h2}$: Co-occurrence Frequency \\
				2 & $F_{m1}$: Number of User & 55 & $F_{m6}$: Fraction of Retweet \\
				3 & $F_{m8}$: Fraction of URL & 56 & $F_{n6}$: D-3 of Exposure vector \\
				4 & $F_{m7}$: Number of Tweet & 57 & $F_{t2}$: mnD \\
				5 & $F_{h3}$: Clarity & 58 & $F_{c4}$: D-4 of Topic Vector \\
				6 & $F_{m2}$: Fraction of User & 59 & $F_{c4}$: D-6 of Topic Vector \\
				7 & $F_{c4}$: D-15 of Topic Vector & 60 & $F_{c1}$: Count of Emoticons  \\
				8 & $F_{c4}$: D-17 of Topic Vector & 61 & $F_{h4}$: Extension Clarity \\
				9 & $F_{m4}$: Fraction of @ & 62 & $F_{n6}$: D-4 of Exposure vector \\
				10 & $F_{u1}$: Activity of User & 63 & $F_{n2}$: Graph Density \\
				11 & $F_{n7}$: Component Fraction & 64 & $F_{n6}$: D-10 of Exposure vector \\
				12 & $F_{c4}$: D-5 of Topic Vector & 65 & $F_{c4}$: D-17 of Topic Vector \\
				13 & $F_{c4}$: D-12 of Topic Vector & 66 & $F_{u3}$: Average of Followers \\
				14 & $F_{c4}$: D-16 of Topic Vector & 67 & $F_{n8}$: Edge Weight	 \\
				15 & $F_{u2}$: Max of Followers & 68 & $F_{c3}$: Sentiment of Hashtag \\
				
				\bottomrule
				
			\end{tabular}
		
		\end{minipage} 
		\vspace{-3mm}
	\end{figure}

	We first evaluate with the whole \textbf{half-year training data}. The fifteen most efficient features (Rank1-Rank15) and the fifteen least efficient features (Rank54-Rank68) are listed in Table~\ref{tab:Feature Rank}. We observe that, {\bf first}, meme features are more efficient,
	$F_{m1}$ and $F_{m7}$ can reflect the present incidence of the topics. 
	A tweet with URL contains more abundant contents that people are more interested in, so $F_{m8}$ is also an efficient feature. 
	{\bf Second}, border users can decide how widely the topic will be disseminated, so $F_{n5}$ is the most efficient feature. 
	{\bf Third}, there are 5 dimensions of Topic Vector among the Top 15, which reflects an actual phenomenon that some specific topics have a relatively high possibility to be popular.
	{\bf Fourth}, there is no Time Series Feature in Top 15. 
	{\bf Fifth}, $F_{n6}$ does not perform as efficient as expected. It is likely that the low dimensions of the vector are too undistinguished to help classify the popularity. While the high dimensions of the vector need a relatively complete dataset to work efficiently. 
	{\bf Sixth}, except for the $F_{c4}$, most of the efficient features are temporal and graphic. 
	However, the complexities of them are commonly high.	
	In the evaluation with only \textbf{7-day} training data, the results are similar. We also find that time series features are more suitable in short period training to describe the latest trends. 
	
	\vspace{-2mm}
	\section{Conclusion}\label{section:conclusion}
	\vspace{-2mm}
	This work proposes a rigorous evaluation scheme for forecasting the rise of certain topics in social media and surveys the existing methods based on it. Prior methods' experiments are based on different datasets, which results in low comparability among them and there is no prior research performing the evaluation. In our scheme, we first classify the methods into two categories: \textit{Feature Oriented Methods} and \textit{Relation Based Methods}. We then perform qualitative evaluation and quantitative experiment based on a real-world dataset. We give out the final ranking values to greatly reflect their comprehensive performance under given scenarios. The evaluation results show that \textit{Feature Oriented Methods} are more suitable for accuracy dominant scenarios and \textit{Relation Based Methods} have better consistency. It also provides several detailed findings. First, denoising essentially suits models to different scenarios. Second, the number of features and data samples have no special relation to the prediction performance. Therefore, choosing efficient features is important. Third, \textit{Feature Oriented Methods} have a higher accuracy but they suffer from heavy overhead such as feature extracting.
	In addition, we evaluate the efficiency of 34 features used in \textit{Feature Oriented Methods}. The results show that features about graph and temporal information are more efficient. In all, our work reveals different characteristics of the methods, provide guidelines on selecting appropriate methods under different scenarios, and help select efficient features for applications.


\begin{thebibliography}{25}


  \bibitem{wu2017sequential} Wu, B. and et al.: Sequential Prediction of Social Media Popularity with Deep Temporal Context Networks.
 In: IJCAI, pp. 3062--3068 (2017)

\bibitem{kong2014realtime} Kong, S. and et al.: On the Real-time Prediction Problems of Bursting Hashtags in Twitter.
 Comput. Sci. 109(3), 246--248 (2014)
 
  \bibitem{Villi2016Recommend} Villi, M. and et al.: Recommend, Tweet, Share: User-Distributed Content (UDC) and the Convergence of News Media and Social Networks. Springer Berlin Heidelberg (2016)

   \bibitem{risk} Talbot, J.: What's right with risk matrices?. Jakeman Business Solutions
  
  \bibitem{ma2013predicting} Ma, Z. and Sun, A. and Cong, G.: On predicting the popularity of newly emerging hashtags in twitter.
  J. Am. Soc. Inf. Sci. Tec. 64(7), 1399--1410 (2013)
 
 \bibitem{kong2014predicting}  Kong, S. and Mei, Q. and Feng, L. and Ye, F. and Zhao, Z.: Predicting bursts and popularity of hashtags in real-time. 
 In: SIGIR, pp. 927--930 (2014)

 \bibitem{cui2012discover} Cui, A. and Zhang, M. and Liu, Y. and Ma, S. and Zhang, K.: Discover breaking events with popular hashtags in twitter.
  In: CIKM, pp. 1794--1798 (2012)

  \bibitem{zhu2014prerecognition} Zhu, T. and Yu, J.: A Prerecognition Model for Hot Topic Discovery Based on Microblogging Data. 
Sci. W. J. 2014, 360934 (2014)

 \bibitem{deng2015effectively} Deng, Z. and et al.: Effectively predicting whether and when a topic will become prevalent in a social network. 
 In: AAAI, pp. 210--216 (2015)

 \bibitem{zhang2014popularity} Zhang, X. and Li, Z. and Chao, W. and Xia, J.: Popularity Prediction of Burst Event in Microblogging. 
 In: WAIM, pp. 484--487 (2014)
 
  \bibitem{thelwall2011sentiment} Thelwall, M. and Buckley, K. and Paltoglou, G.: Sentiment in Twitter events.
  J. Am. Soc. Inf. Sci. Tec. 62(2), 406--418 (2011)

 \bibitem{li2012twevent} Li, C. and Sun, A. and Datta, A.: Twevent: segment-based event detection from tweets. 
 In: CIKM, pp. 155--164 (2012)


 
  \bibitem{chen2014non} Chen, F. and et al.: Non-parametric scan statistics for event detection and forecasting in heterogeneous social media graphs.
  In: SIGKDD, pp. 1166--1175 (2014) 




 \bibitem{gao2014effective} Gao, S. and Ma, J. and Chen, Z.: Effective and effortless features for popularity prediction in microblogging network. 
 In: WWW, pp. 269--270 (2014)
 
 \bibitem{yin2011geographical} Yin, Z. and Cao, L. and Han, J. and Zhai, C. and Huang, T.: Geographical topic discovery and comparison. 
 In: WWW, pp. 247--256 (2011)

 \bibitem{sakaki2010earthquake} Sakaki, T. and Okazaki, M. and Matsuo, Y.: Earthquake shakes Twitter users: real-time event detection by social sensors. 
 In: WWW, pp. 851--860 (2010)


 \bibitem{amodeo2011hybrid} Cataldi, Amodeo, G. and Blanco, R. and Brefeld, U.: Hybrid models for future event prediction. 
 In: CIKM, pp. 1981--1984 (2011)

 \bibitem{du2011microblog} Du, Y. and He, Y. and Tian, Y. and Chen, Q. and Lin, L.: Microblog bursty topic detection based on user relationship. 
 In: ITAIC, pp. 260--263 (2011)

 \bibitem{ishikawa2012hot} Ishikawa, S. and Arakawa, Y. and Tagashira, S. and Fukuda, A.: Hot topic detection in local areas using Twitter and Wikipedia. 
 In: ARCS, pp. 1--5 (2012)

 \bibitem{chen2013emerging} Chen, Y. and Amiri, H. and Li, Z. and Chua, T.: Emerging topic detection for organizations from microblogs. 
 In: SIGIR, pp. 43--52 (2013)

 \bibitem{becker2011beyond} Becker, H. and Naaman, M. and Gravano, L.: Beyond Trending Topics: Real-World Event Identification on Twitter.
 In: ICWSM, pp. 11: 438--11: 441 (2011)
 
\bibitem{zhang2015event} Zhang, X. and et al.: Event detection and popularity prediction in microblogging.
 Neurocomputing 129, 1469--1480 (2015)
 
  \bibitem{wang2017predicting}  Wang, X. and Wang, C. and Ding, Z. and Zhu, M. and Huang, J.: Beyond Trending Topics: Real-World Event Identification on Twitter.
 J. Intell. Inf. Syst.  (2017)










 \bibitem{Page1998The} Page, L. and Brin, S. and Motwani, R. and Winograd, T.: The PageRank citation ranking: bringing order to the web. 
 Stanford InfoLab (1999)

 \bibitem{chen2013latent} Chen, G. H. and Nikolov, S. and Shah, D.: A Latent Source Model for Nonparametric Time Series Classification.
 In: NIPS - Volume 1, pp. 1088--1096 (2013)

 
 \bibitem{matsubara2017nonlinear} Kong, S. and Mei, Q. and Feng, L. and Zhao, Z. and Ye, F.: Nonlinear Dynamics of Information Diffusion in Social Networks.
 ACM Trans. Web 11(2), 1--40 (2017)

 \bibitem{hoang2017gpop} Hoang, M. X and et al.: GPOP: Scalable Group-level Popularity Prediction for Online Content in Social Networks.
 In: WWW, pp. 725--733 (2017)


\bibitem{liu2014trending} Liu, Y. and Han, W. and Tian, Y. and Que, X. and Wang, W.: Trending topic prediction on social network.
 In:  IC-BNMT, pp. 149--154 (2013)

\end{thebibliography}
\end{document}